# Linear Optical Modulators for Prospective Communications at the 2 μm Waveband

Jia Xu Brian Sia, Xiang Li, X. Guo, Jiawei Wang, Wanjun Wang, Zhongliang Qiao, Callum G. Littlejohns, Chongyang Liu, Kian Siong Ang, Graham T. Reed, and Hong Wang

**Abstract**— **The 2 μm waveband is an area that could have significant technological consequences, with applications ranging from spectroscopy, LIDAR and free-space communications. The development of the thulium-doped fiber amplifier, hollow-core photonic bandgap fiber and 2 μm GaSb-based diode lasers has highlighted the ability of the waveband in alleviating the fiber capacity crisis in the incumbent communication infrastructure. The above has initiated vibrant development in the silicon photonic-space at 2 μm, where the area is capable of enabling highly-integrated photonic circuits, and potentially at low-cost and high-volumes. However, as of now, modulator linearity at 2 μm has not been addressed. The metric, as characterized by spurious free dynamic range is imperative for numerous applications such as RF photonic links for 5G and digital analog transmission in coherent communications. The development of linear optical modulators will be crucial in bringing these applications to the 2 μm. In view of that, this work is the first to address modulator linearity at the 2 μm, where the ring-assisted Mach-Zehnder modulator is developed, indicating spurious free dynamic range as high as 95 dB.Hz$^{2/3}$. It is found that that modulator spurious free dynamic range has a strong dependence on modulator bias voltage where it impacts the linearity of the transfer function in which the input RF signal is applied upon. The demonstrated modulator indicates favorable performance within silicon photonic modulators developed at 2 μm with bandwidth exceeding 17.5 GHz and modulation efficiency ranging from 0.70 to 1.25 V.cm.**

## I. INTRODUCTION

The fiber optic infrastructure at C-band, currently, forms the fundamental basis of the internet, enabling low-latency, high-capacity transmission of information [1]. The increased demand for information capacity has been unabating, driven by technologies, such as the transition to 5G, the increased uptake in video conferencing and popularity of high-resolution video streaming services [2]. In view of the growing capacity requirements, current approaches have involved: (1) increasing the bandwidth of modulators [3], (2) wavelength-division multiplexing [4], (3) advanced modulations formats [5]. However, it is imperative to note, as highlighted by the Shannon limit, there is an upper limit to the number of bits and equivalently, energy that can be carried by the single-mode fiber [6]. The current consensus is that the Shannon limit will be breached in the near future and there is an urgent need to address the challenge [7].

In view of the limitation, it is important for the development of a solution that will enable a sustainable pathway towards overcoming the capacity limit in the long run. A potential candidate towards overcoming this difficulty is the possibility of optical communications at the 2 μm waveband [8]. The development of 2 μm GaSb-based diode lasers [9], high-capacity, low-loss, hollow-core photonic bandgap fiber [10] and thulium-doped fiber amplifiers with high small-signal and low-noise figures [11] have addressed the infrastructural aspect of the wavelength region. Silicon photonics, due to its compatibility with mature silicon manufacturing, is commensurate with low-cost, high-volume production [12]. Driven the possibility of 2 μm optical communications, there has been vibrant developments in the area: hybrid silicon lasers [13]-[16], photodetectors [17]-[19], passive components [20]-[22] as well as modulators [23]-[26]. This trend is further supplemented by the stronger plasma dispersion effect [23] and significantly lower two-photon absorption [27] in contrast to the C-band. The latter is proving to be a formidable challenge in industry. Of late, the development of modulators operating at 2 μm have been promising, with microring and Mach-Zehnder modulators reaching bandwidths of 18 GHz [23]-[24], demonstrating data transmission rates of up to 80 Gbits/s (PAM-4). However, it is known that modulator linearity is a salient metric for numerous applications, such as digital-to-analog (DAC) transmission for coherent communications as well as RF photonic links for 5G and free-space optical communications [28]-[29]. The development of a linear silicon photonic optical modulator will extend the mentioned

J. X. B. Sia, X. Guo, J. Wang, W. Wang, H. Wang are with the School of Electrical and Electronic Engineering, Nanyang Technological University, 50 Nanyang Avenue, Singapore 639798. (jiaxubrian.sia@ntu.edu.sg, ewanghong@ntu.edu.sg)

K. S. Ang, is with CompoundTek, 5 International Business Park, Singapore 609914.

X. Li, and C. Liu are with the Temasek Laboratories, Nanyang Technological University, 50 Nanyang Avenue, Singapore 639798.
C. G. Littlejohns and G. T. Reed are with the Optoelectronics Research Centre, Unviersity of Southampton, Southampton, SO17 1BJ, UK
Z. Qiao, is with the Key Laboratory of Laser Technology and Optoelectronic Functional Materials of Hainan Province, and School of Physics and Electronic Engineering, Hainan Normal University, Haikou 571158, China.
.

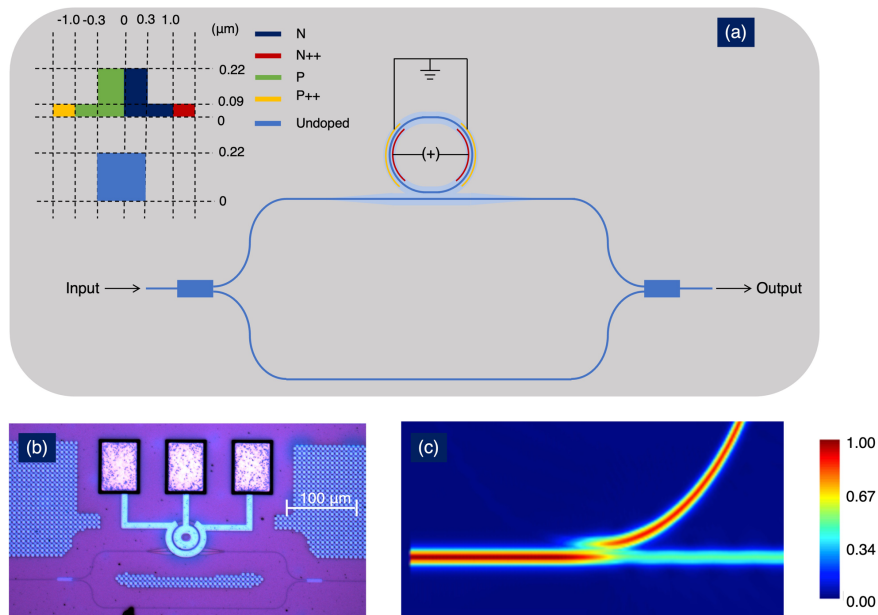

Fig. 1. (a) 2-D schematic of the RAMZM. The strip waveguide and slab are indicated in blue and light blue respectively. The cross-sectional schematic of the strip and doped rib waveguide are illustrated at the inset. (b) Micrograph image of the RAMZM, with the associated scale-bars. (c) Top-view electric-field distribution of the coupling region between the racetrack phase shifter and one of the Mach-Zehnder arms; the power coupling coefficient is 0.77.

applications to the 2 μm wavelength region, significantly alleviating the capacity concerns at the incumbent wavelength regions (C-band). The linearity of a modulator can be defined via the spurious free dynamic range (SFDR). One can define SFDR as the ratio between the signal of interest and the most significant dynamic spurious signal in the background. In other words, it refers to the operational dynamic range of a modulator prior to the onset of signal distortion.

Pertaining to the current state of modulator development at the 2 μm waveband [23]-[26], it is imperative to notice that modulator linearity, or rather SFDR is an area that has not been addressed. It is of note that the transfer function of the Mach-Zehnder modulators (MZM) is negatively nonlinear [28], [30]-[31]. This, coupled with the third-order nonlinearity of silicon index change will impact the SFDR of silicon-based modulators. While linearization through external electronics can be a solution, increased complexity in system complexity will result, which could be more prevalent given the lower maturity of the 2 μm waveband [28]. It is important for the development of 2 μm modulators with minimal compromise in system complexity to enable potential applications in DAC-based coherent communications and RF photonic links, as mentioned above [28]-[29]. This could alleviate the upcoming capacity crisis in the incumbent fiber infrastructure.

In this work, we address the issue of linearity at 2 μm through the development of ring-assisted MZM (RAMZM) operating at 2 μm for the first time. Through the integration of a racetrack phase shifter with a Mach-Zehnder interferometer (MZI), linear modulation can be realized through the reduction of the negative nonlinearity associated with the MZI. A static extinction ratio of 21.5 dB, with a modulation efficiency ($V_\pi$ $L_\pi$) ranging from 0.70 to 1.25 V.cm, subject to bias voltages ($V_{bias}$) of 1 to 8 v is demonstrated. The bandwidth of the RAMZM is measured to exceed that of the 2 μm characterization system at 17.5 GHz, comparing favorably with the current work involving silicon photonic modulators at the 2 μm waveband [23]-[26]. SFDR as high as 95 dB.Hz$^{2/3}$ is demonstrated. In addition, the linearity (SFDR) of the 2 μm RAMZM is studied as a function of $V_{bias}$ where a significant $V_{bias}$ dependence is observed.

## II. Modulator Design and Operation Principle

The 2-D schematic of the RAMZM operating at the 2 μm waveband is shown in Fig. 1(a); micrograph image with the associated scale bar indicated in Fig. 1(b). The modulator is fabricated on the 220 nm silicon-on-insulator platform, based on our developed platform; the platform is available commercially to all, and detailed fabrication steps are elaborated in [32]. The arm length of the MZI is symmetrical at 375 μm, connected by two 1 × 2 multi-mode interferometer (MMI) at the ends. The MMI, designed to operate at 1970 nm (λ = 1970 nm) has a MMI width, length and taper spacing, length of 6, 22.6 μm and 3.14, 10 μm respectively. The input and output mode of the MMI tapers from a width of 1.35 to 0.6 μm, over a taper length of 10 μm to ensure lossless adiabatic transition. The Mach-Zehnder arms consists of two 90º bends on both sides, with radius of 20 μm, thereby eliminating bending losses. Two types of fundamental transverse electric (TE00) waveguide are utilized in the modulator: strip waveguides with width of 0.6 μm, rib waveguides with width of 0.6 μm and rib height of 0.09 μm (Fig. 1(a), inset). The rib waveguide is implemented for the integration of the P-N junction within the racetrack phase shifter where the strip waveguide mode transits to the rib waveguide mode adiabatically when the width of the rib increases to 10 μm over a length of 30 μm. The rest of the modulator is composed of strip waveguides. With regards to the racetrack phase shifter, the P-N junctions, comprises of 44 % of the 180º bends on both sides. For the P-N junction cross-section as illustrated by the inset of Fig. 1(a), two levels of n and p-type doping is implemented about the center of the Si rib waveguide. The P++ and N++ doping are highly doped, forming ohmic contacts. The doping concentration of P and N are about $5 \times 10^{17} \text{cm}^{-3}$. In order to avoid resonant photon absorption by the heavily doped regions, the P++ and N++ regions are placed 1.0 μm away from the center of the rib waveguide; overlap analysis of the cross-sectional waveguide mode indicates that 99 % of the optical energy is contained within the P and N-doped regions. The straight section of the racetrack phase shifter is 2.25 μm-long, with a waveguide edge-to-edge gap of 0.2 μm, designed in view of modulator SFDR performance. This corresponds to a power coupling coefficient of 0.77; the simulated top-view electric-field distribution in the coupling region of the racetrack resonator is shown in Fig. 1(c). The racetrack resonator is designed to be overcoupled in order to reduce the negative nonlinearity of the MZI. One of the key advantages of the RAMZM structure is that unlike microring modulators where critical coupling between the resonator and the bus waveguide is required, for the case of RAMZM, the requirement is for the racetrack phase shifter to be overcoupled, thereby increasing the fabrication tolerance of the modulator.

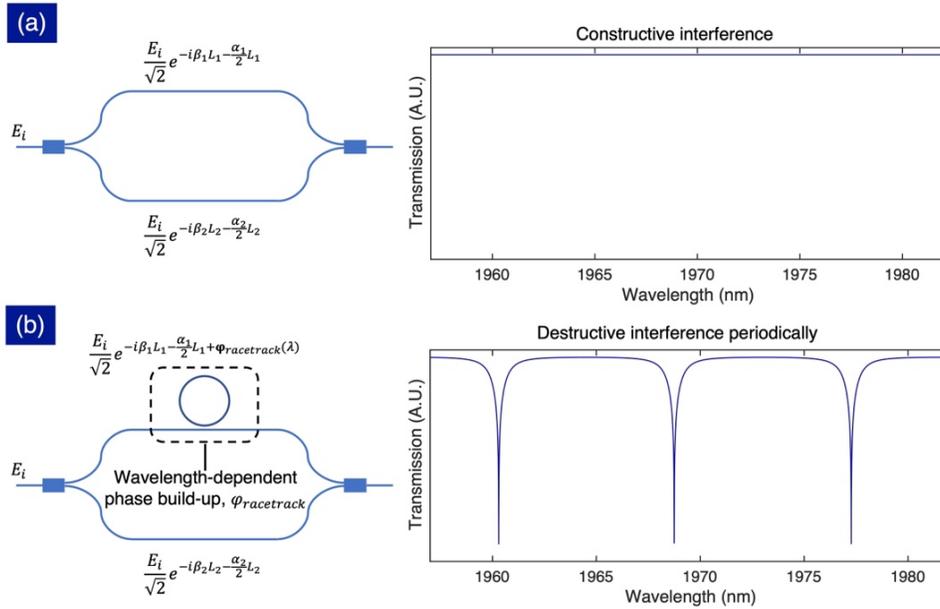

Fig. 2. (a) 2-D illustration of symmetrical MZI (left), illustration of associated optical spectrum (right); $E_i$ is the input field and $\beta$ is the propagation constant ($\beta_1 = \beta_2$), $\alpha$ is the loss coefficient ($\alpha_1 = \alpha_2$) on each arm. (b) 2-D illustration of RAMZM (left), illustration of associated optical spectrum; $E_i$ is the input field and $\beta$ is the propagation constant ($\beta_1 = \beta_2$), $\alpha$ is the loss coefficient on each arm. The $\varphi$ racetrack is wavelength-dependent phase build-up by racetrack phase shifter.

Similar to MRM and MZM, the RAMZM is an intensity modulator. In order to understand the operating principle of RAMZM, a symmetrical MZI can be evaluated first (Fig. 2(a)) where the arm lengths are equivalent. Light is split equally into the two Mach-Zehnder arms of the same length. As the optical path length and, subsequently, the phase of light in the two Mach-Zehnder arms are equivalent, the lightwave at the two paths will interfere constructively, resulting in the spectrum illustrated in Fig. 2(a). On the other hand, for a RAMZM, where a racetrack resonator is integrated on one of the arms of the symmetrical MZI (Fig. 2(b)), when the phase shift build-up by the racetrack ($\varphi_{racetrack}$) resonator at a wavelength corresponds to odd multiples of $\pi$ ($\varphi_{racetrack} = \pi \times$ odd number), the lightwave along the two Mach-Zehnder arms would interfere destructively and correspond to a minimum in the optical spectrum; this condition is fulfilled periodically across an optical spectrum and as such, this results in the formation of periodic resonances (Fig. 2(b)). As such, one will be able to shift the resonant wavelength of the RAMZM structure through the integration of P-N junctions within the racetrack phase shifter and subsequently vary the voltage applied.

## III. Static Characterization and Bandwidth

### A. Transmission, Transfer Function and $V_\pi L_\pi$

The static characteristics of the RAMZM is measured via the following experimental setup (Fig. 3). An external wavelength-tunable laser source operating at 2 μm is coupled to the fiber polarization controller (FPC), in and out of the RAMZM via cleaved vertical coupling fibers angled at 10º, to the photodetector. Grating couplers designed to operate at the 2 μm waveband are implemented as I/O for the RAMZM (period = 0.93 μm, duty cycle = 0.55, etch depth = 70 nm). DC bias voltage ($V_{bias}$) is applied to the racetrack phase shifter via a DC probe.

Fig. 4(a) shows the normalized transmittance spectra of the RAMZM when the $V_{bias}$ applied to the racetrack phase shifter is 0 v. The transmittance spectra is normalized against a straight waveguide with the abovementioned grating coupler as I/O. As expected, periodic resonances with free-spectral range of 7.7 nm is measured. For the subsequent measurements in this work, $\lambda \approx 1970$ nm; where the extinction ratio is measured to be 21.5 dB. The transfer function of the RAMZM is illustrated in Fig. 4(b). A linear operating region can be observed when $V_{bias}$ ranges from 1 to 7 v; the RAMZM operates in the depletion-mode. In the subsequent section, measurement results indicates that the varied application of $V_{bias}$ will have significant impact on the SFDR of the RAMZM. Fig. 4(d) indicates the zoomed-in transmission spectra of the RAMZM when $\lambda \approx 1970$ nm, to indicate the wavelength shift as a function of $V_{bias}$. When $V_{bias}$ increases, it can be seen that the resonant wavelength shifts towards longer wavelength. According to expectation, increasing $V_{bias}$ in the negative polarity reduces the refractive index of the waveguide, resulting in the redshift of the resonant wavelength as the condition ($\varphi_{racetrack} = \pi \times odd\ number$) is fulfilled at shorter wavelengths. The $V_\pi L_\pi$ of the RAMZM is shown in Fig. 4(d), where it increases from a value of 0.70 to 1.25 V.cm. The degradation of $V_\pi L_\pi$ is attributed to the sublinear relationship between the depletion width ($W_{P-N}$) of the P-N junction and $V_{bias}$ ($W_{P-N} \alpha \sqrt{V_o - V_{bias}}$) [33].

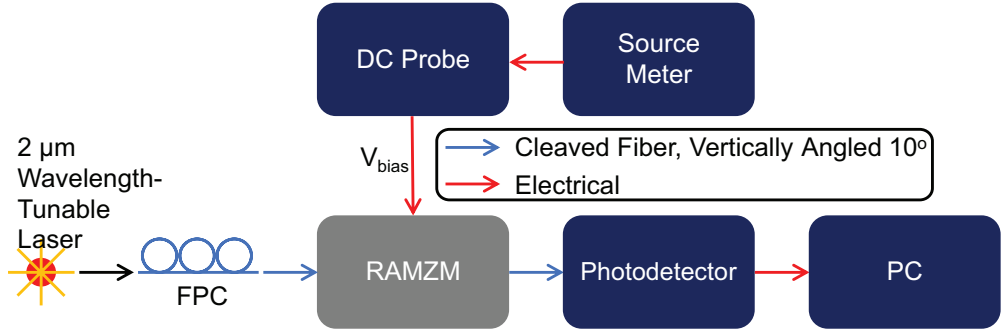

Fig. 3. Experimental setup for static characterization: transmission, transfer function, $V_\pi L_\pi$.

### B. Bandwidth

The electro-optic response of the RAMZM is characterized via the experimental setup illustrated in Fig. 5. For each value of $V_{bias}$ that was measured, the wavelength of the tunable laser source is tuned 3-dB down to the resonance near $\lambda \approx 1970$ nm, through the FPC and coupled in and out of the RAMZM via grating couplers through vertically angled ($10^o$) cleaved fibers. A RF input from the lightwave component analyzer (LCA) and $V_{bias} = 0, 1, 3, 5, 7, 8$ v are applied to a bias tee, sent to the RAMZM via the RF probe. The output fiber is connected to the high-speed photodetector and then the LCA, where the electro-optic response of the modulator is measured. It is of note that the measurement setup is limited by the bandwidth of the high-speed photodetector, with a bandwidth to 17.5 GHz.

Fig. 6 shows the measured electro-optic response of the RAMZM when $V_{bias} = 1, 3$ v. The bandwidth is measured to be 16.2 GHz when $V_{bias} = 1$ v. However, when $V_{bias}$ is increased further, to 3 v, it is found that the bandwidth of the measurement setup limits characterization. The increase in bandwidth with increase in $V_{bias}$ is attributed to the widening of the junction depletion width of the phase shifter, leading to a drop in junction capacitance and subsequently, RC time constant [33]. It can be concluded with certainty that the bandwidth of the RAMZI exceeds 17.5 GHz when $V_{bias}$ is larger than 3 v; the bandwidth of the high-speed photodetector in the measurement setup is 17.5 GHz. As seen in Fig. 6, optical peaking, facilitated by the intrinsic time dynamics of the racetrack phase shifter can be observed, resulting in the extension of the electro-optic bandwidth of the modulator [34]; optical peaking is all observed at $V_{bias} = 1, 3, 5, 7$ v.

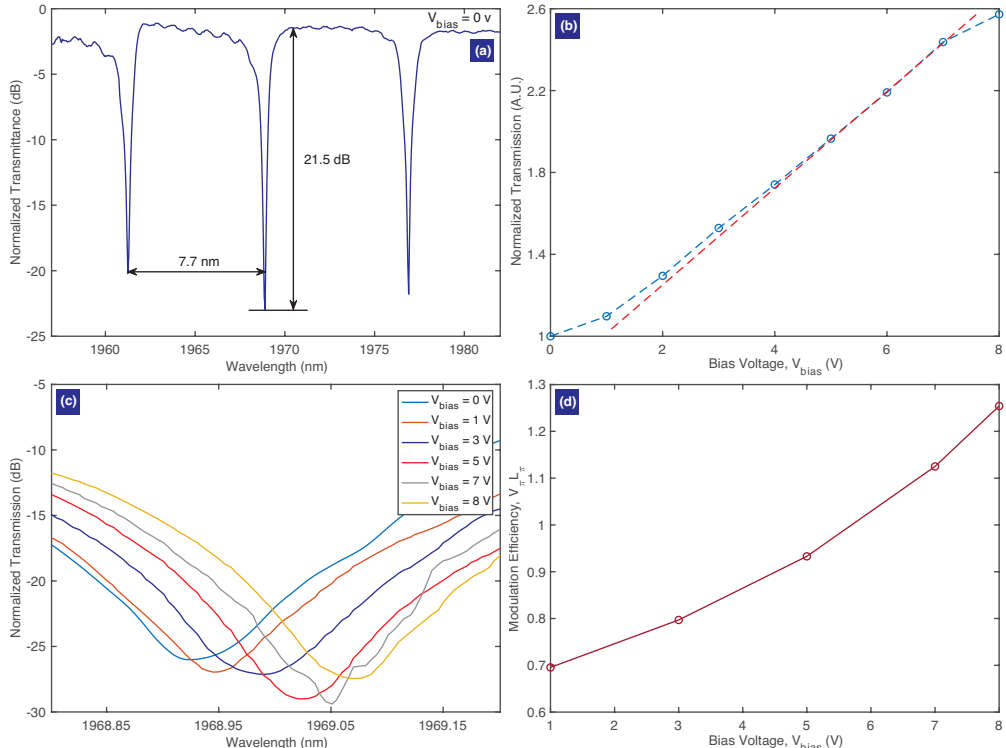

Fig. 4. (a) Transmittance spectra of the RAMZM when $V_{bias} = 0$ v. (b) Transfer function of the RAMZM. (c) Zoomed-in resonant shift when $V_{bias} = 0, 1, 3, 5, 7, 8$ v. (d) $V_\pi L_\pi$ of the RAMZM.

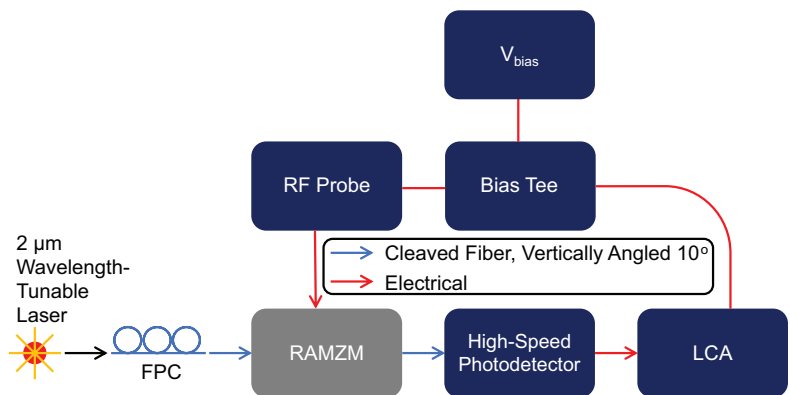

Fig. 5. Experimental setup for RAMZM bandwidth characterization.

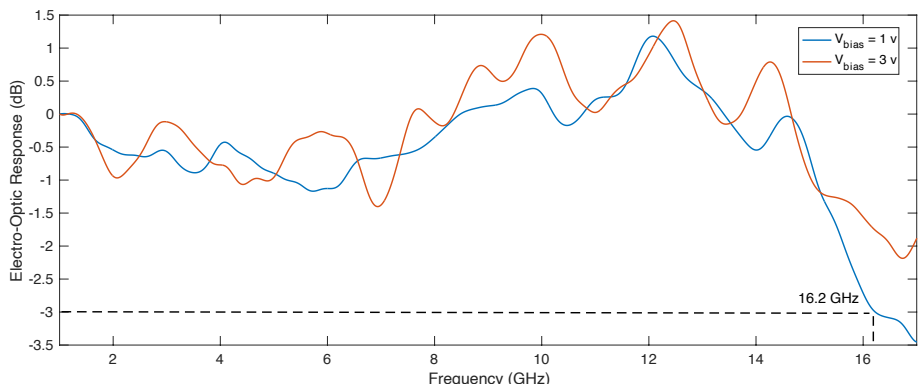

Fig. 6. Measured electro-optic response when V_bias = 1, 3 v. Our measurement setup is limited by the bandwidth of the high-speed photodiode, at 17.5 GHz.

## IV. SFDR

The RAMZM is modulated along its transfer function during operation, where a schematic indicating the transmission of the modulator against V_bias as such is illustrated in Fig. 7(a); the measured transfer function of the RAMZM is shown in Fig. 4(b). Shown in Fig. 7(a), a modulating voltage signal (input RF signal) is applied to the RAMZM, resulting in an output modulated light signal (output RF signal). As it can be seen in Fig. 7(a), the nonlinearity of the transfer function introduces nonlinear distortions to the output signal of the modulator. However, it is important to note that in most analog applications, the operating bandwidth is lower than one octave, as such, third-order intermodulation distortion (IMD) is required to be addressed [35]. The IMD3 can be observed at $2f_1-f_2$ and $2f_2-f_1$ in the RF spectrum when two tones $f_1$ and $f_2$ are applied to the modulator as illustrated in Fig. 7(b). The SFDR is obtained through the relative ratio between the fundamental tone $(f_1, f_2)$ and the strongest spurious signal in the output. It can be defined as the difference in output RF power between the fundamental tone and IMD3 when the power level of IMD3 reaches the noise floor of the system.

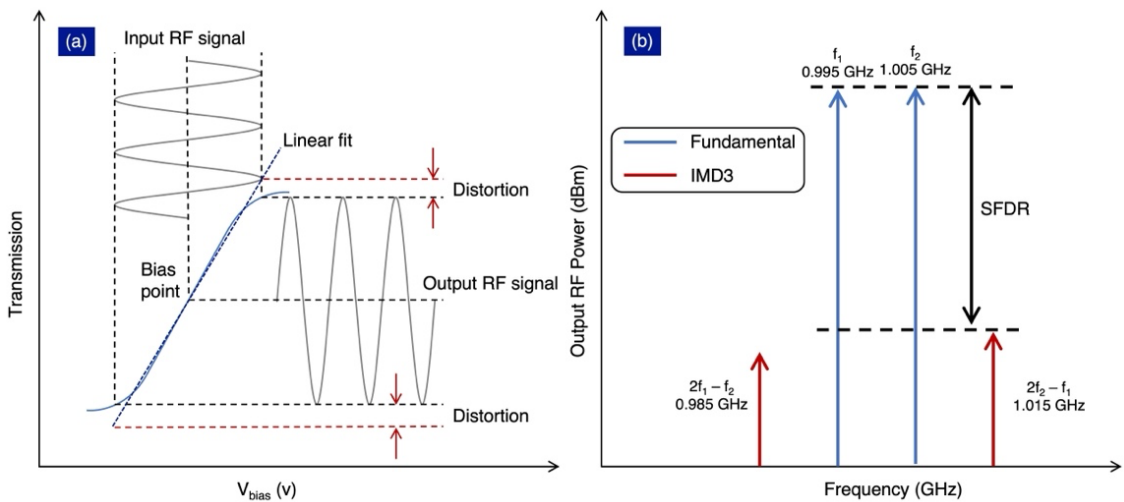

Fig. 7. (a) Modulator transfer function, referring to transmission against Vbias. Illustration of how nonlinear distortion is formed because of the nonlinearity of the transfer function. (b) illustration of fundamental (F), third-order intermodulation distortion (IMD3), and spurious free dynamic range (SFDR) in the RF spectrum.

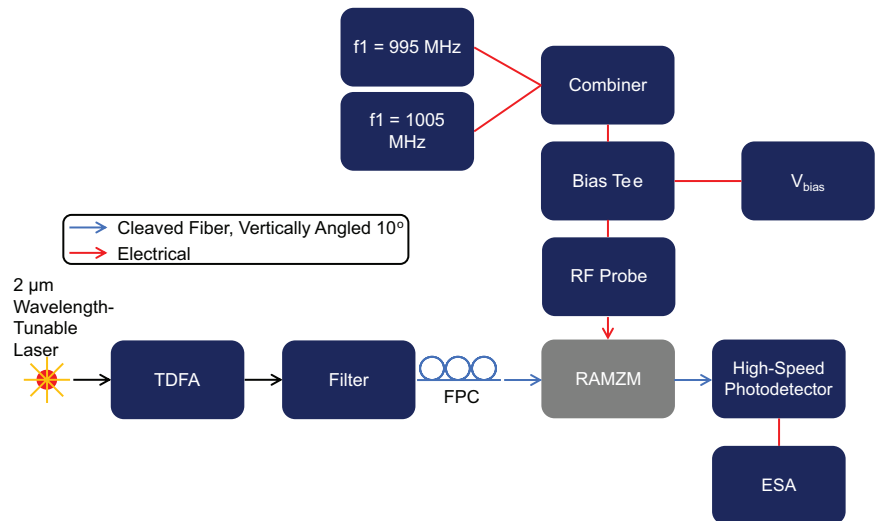

Fig. 8. Experimental setup for SFDR measurement. spectrum.

The experimental setup involved to measure the SFDR of the RAMZM is illustrated in Fig. 8, where the output of a 2 μm external wavelength-tunable laser source is coupled to the RAMZM through a FPC and then a thulium-doped fiber amplifier (TDFA). The resulting source is coupled to the modulator. Fundamental tones, $f_1$ (0.995 GHz) and $f_2$ (1.005 GHz), are combined and applied to the RAMZM; the input power for the RF tones are varied for SFDR measurement in Fig. 9, 10 and 11. The output optical signal from the modulator is sent to a high-speed photodetector, and then, the electrical spectrum analyzer (ESA). The noise level of the measurement setup was measured to be -160 dBm, when the resolution bandwidth used in the ESA is 1 Hz.

SFDR measurements of the RAMZM when $V_{bias} = 1, 3, 5, 7$ v are shown in Fig. 9(a)-(d) respectively. As an example, the evolution of the RF spectrum at input tone frequencies of 0.995, 1.005, 0.985, 1.015 GHz against input RF power is indicated in (a)-(d) respectively for Fig. 10, 11, where $V_{bias} = 5, 7$ v. Due to significant data acquisition times at resolution bandwidth of 1 Hz, the spectrum collection in Fig. 10, 11 are performed over a span to 10 kHz where no spurious noise is observed in the vicinity of the measured spectrum except at IMD3 (985, 1015 MHz); this is confirmed with resolution bandwidths of 1 MHz when data acquisition times are significantly shorter ,but do not enable the observation of fine spectral features. From Fig. 9(a)-(d), SFDR is observed to have significant $V_{bias}$ dependence, where SFDR is found to be 91 and 95 dB.Hz$^{2/3}$ when $V_{bias} = 3,$

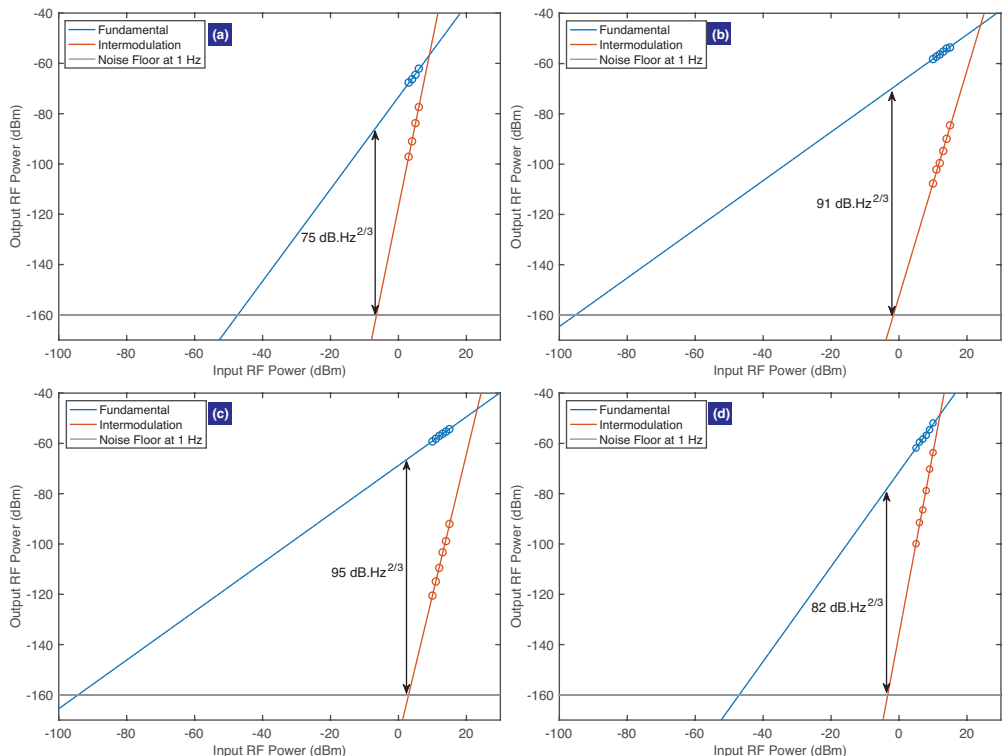

Fig. 9. SFDR of the RAMZM when, (a) $V_{bias} = 1$ v, (b) $V_{bias} = 3$ v, (b) $V_{bias} = 5$ v, (d) $V_{bias} = 7$ v.

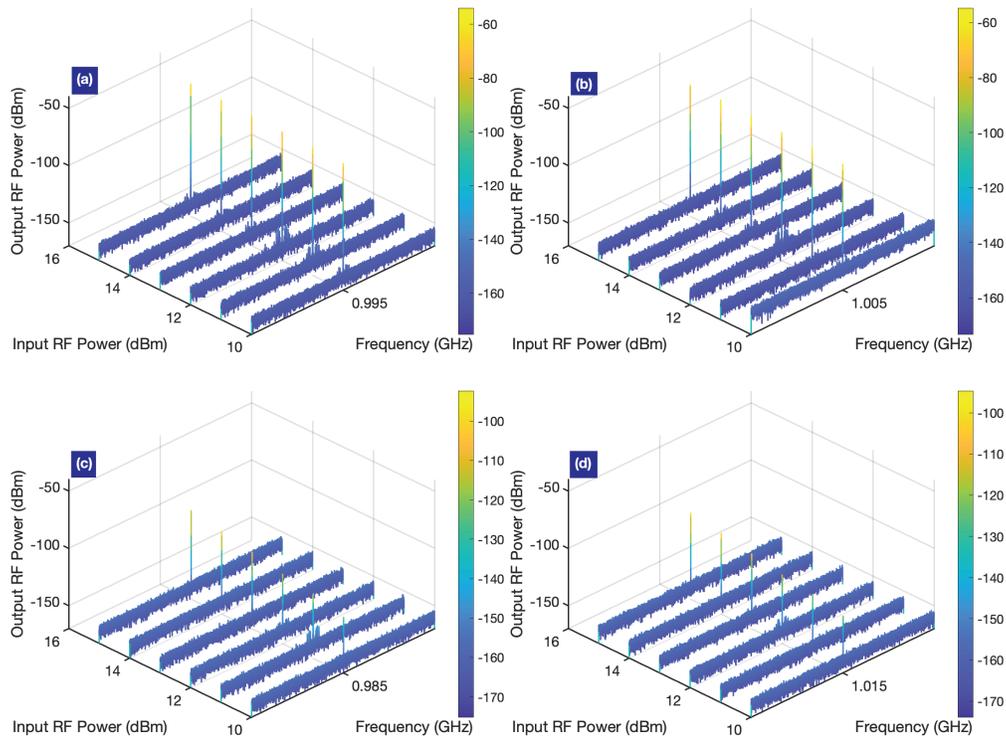

Fig. 10. RF spectrum at (a) 0.995, (b) 1.005, (c) 0.985, (d) 1.015 MHz, when $V_{bias}$ = 5 v.

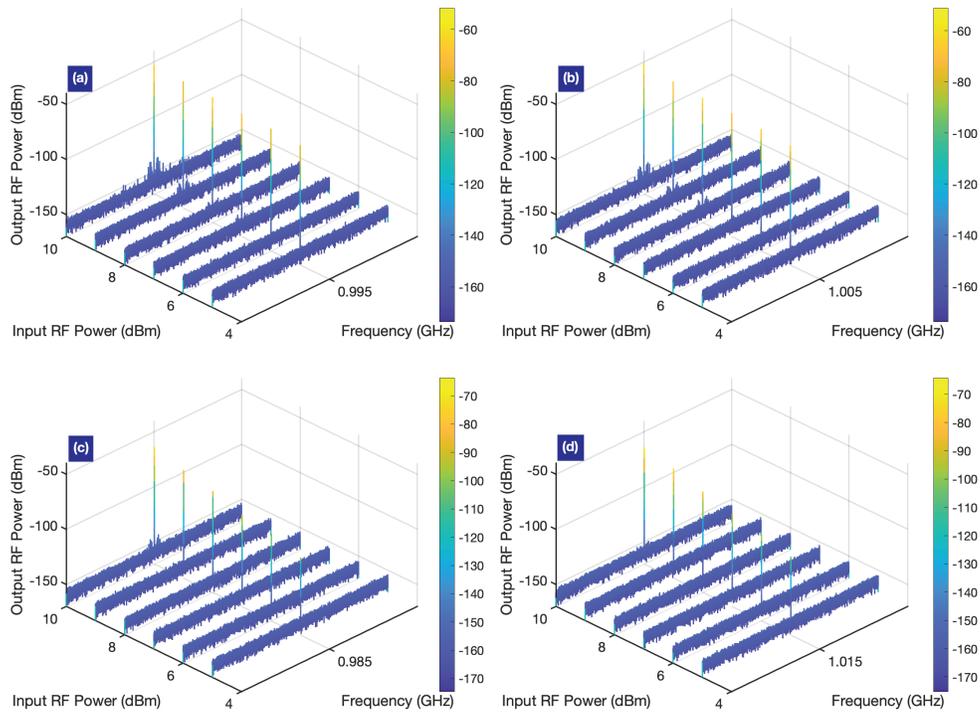

Fig. 11. RF spectrum at (a) 0.995, (b) 1.005, (c) 0.985, (d) 1.015 MHz, when $V_{bias}$ = 7 v.

5 v respectively (Fig. 9(b), (c)). On the other hand, degradation in SFDR is seen when $V_{bias}$ = 1, 7 v (Fig. 9(a), (d)). This can be explained by the transfer function of the RAMZM indicated in Fig. 4(b) where there is high linearity at $V_{bias}$ = 3, 5 v, and subsequently, lower linearity at the $V_{bias}$ = 1, 7 v. It can be concluded that modulator SFDR is $V_{bias}$-dependent due to the linearity of the transfer function corresponding to $V_{bias}$. In addition, when $V_{bias}$ = 1, 7 v, the SFDR at $V_{bias}$ = 1 v is lower as the P-N junction is operating closer to the positive polarity, operating in the carrier injection mode. The carrier injection has a strong nonlinearity, resulting in a further degradation in SFDR [36].

## V. CONCLUSION

The 2 μm waveband is of much intrigue technologically, where the wavelength region is capable of lending itself to a wide range of applications. More specifically, the development of 2 μm GaSb-based diode lasers [9], high-performance thulium-doped fiber amplifier [10] and hollow-core photonic bandgap fiber [11] has highlighted the potential of the waveband in mitigating capacity concerns at the incumbent communication infrastructure. Modulator linearity is a salient metric amongst optical modulators in enabling applications such as DAC-based coherent communications and RF photonic links where this works addresses this issue for the first time at the 2 μm waveband [28]-[29]. Through the development of the RAMZM operating at 2 μm, SFDR as high as 95 dB.Hz$^{2/3}$ is measured. It is found that modulator linearity has a significant dependence on $V_{bias}$, where $V_{bias}$ affects the linearity of the transfer function that the input RF signal modulates upon. In addition, the RAMZM indicates favorable performance against the current state of 2 μm silicon photonic modulators where bandwidth exceeding 17.5 GHz and $V_\pi L_\pi$ ranging from 0.70 to 1.25 V.cm is characterized.

## ACKNOWLEDGMENT

The authors would like to acknowledge the support of the Ministry of Education (MOE), Singapore (MOE-T2EP50121-0005). Partial acknowledgement is given to the National Natural Science Foundation of China (62274048).

Author's biographies not available at time of publication,